# Theory on the mechanisms of combinatorial binding of transcription factors with DNA

*R. Murugan*[*]
*Department of Biotechnology, Indian Institute of Technology Madras*
*Chennai, India*

[*] rmurugan@gmail.com

**Abstract**
We develop a theoretical framework on the mechanism of combinatorial binding of transcription factors (TFs) with their specific binding sites on DNA. We consider three possible mechanisms viz. monomer, hetero-oligomer and coordinated recruitment pathways. In the monomer pathway, combinatorial TFs search for their targets in an independent manner and the protein-protein interactions among them will be insignificant. The protein-protein interactions are very strong so that the hetero-oligomer complex of TFs as a whole searches for the cognate sites in case of hetero-oligomer pathway. The TF which arrived first will recruit the adjacent TFs in a sequential manner in the recruitment pathway. The free energy released from the protein-protein interactions among TFs will be in turn utilized to stabilize the TFs-DNA complex. Such coordinated binding of TFs in fact emerges as the cooperative effect. Monomer and hetero-oligomer pathways are efficient only when few TFs are involved in the combinatorial regulation. Detailed random walk simulations suggest that when the number of TFs in a combination increases then the searching efficiency of TFs in these pathways decreases with the increasing number of TFs in a power law manner. The power law exponent associated with the monomer pathway seems to be strongly dependent on the number of TFs, distance between the initial position of TFs from their specific binding sites and the hop size associated with the dynamics of TFs on DNA. Here hop size is positively correlated with the degree of condensation or supercoiling of DNA. The searching efficiency of TFs in the coordinated recruitment pathway with propagating cooperative effects decreases with increasing number of TFs in a logarithmic manner. Since the combinatorial regulation is mandatory for eukaryotic systems to avoid the speed-fidelity issues related to the searching of TFs over large genome size and nuclear volume, the recruitment pathway with cooperative effects is the paramount mechanism for most of the eukaryotic systems.

**Key words**
Combinatorial binding, one dimensional diffusion, random searching, site-specific protein-DNA interactions, transcription regulation

## 1. Introduction

Site-specific interaction of proteins with the genomic DNA in the presence of enormous amount of **n**on-**s**pecific **b**inding **s**ites (**nSBS**) is a vital step in replication, transcription and translation of the genetic information. Transcription factors (**TF**s) play critical roles in the initiation of transcription especially in eukaryotes [1-4]. The TFs associated with the promoter of a gene first bind with the corresponding *cis*-acting regulatory modules (**CRM**s or aka enhancers) of that gene to form a complex (enhancer-transcription factor complex, **ETF**). CRMs are located either upstream or downstream of the promoter sequences of the corresponding genes. This ETF complex then distally acts on and stabilize the already formed RNA polymerase II (RNAPII) - promoter complex (**RP**) which results in the initiation of transcription [1, 3, 5, 6]. In case of eukaryotic systems such as human, higher plants and animals, multiple transcription factors regulate the associated genes in a combinatorial manner with AND logic rather than on a one-to-one basis as in case of prokaryotic systems such as *E. coli* bacteria [1-3].

In human cell, there are $\sim 10^3$ different TFs which regulate $\sim 10^4$ genes in a combinatorial manner [7]. Denote these TFs as $TF_1$, $TF_2$, and so on. The regulated genes are denoted as $G_1$, $G_2$, and so on (**Fig. 1**). Then a combination of TFs such as ($TF_1$, $TF_3$, and $TF_5$) will regulate gene $G_1$ with AND logic via binding at sequentially located binding sites $X_1X_3X_5$ (CRM for $G_1$) and some other combination of TFs such as ($TF_2$, $TF_3$, $TF_4$, and $TF_6$) regulates $G_2$ via binding at $X_2X_3X_4X_6$ (CRM for $G_2$) and so on. Here we denote the binding site for $TF_1$ as $X_1$, $TF_2$ for $X_2$ and so on. Such combinatorial regulation plays critical roles in minimizing the speed-fidelity problem that arises in the site-specific recognition of the specific binding sites (**SBS**) on DNA by the respective TFs [3] particularly for large genome sizes and cell volume as in case of eukaryotes. Since the overall site-specific binding energy is the sum of individual binding energies of all the TFs in a combination originating from both DNA-protein and protein-protein interactions, specific binding energy associated with the individual TF protein may be close to the background non-specific binding energy. This in turn helps TFs to escape from the kinetic traps those are present along the sequence dependent nonspecific DNA-protein interaction-energy landscape which are generated by the randomly occurring analogous binding sites of the genomic DNA [8].

It is generally believed that the binding of multiple TFs exhibits a net cooperative effect through a combination of protein-protein and protein-DNA interactions. This means that binding of a TF with its respective binding site on DNA will be enhanced by a neighboring already bound TF. Binding of multiple TFs at sequentially located SBSs would also incur slowness due to simultaneous random searching problem on which the degree of supercoiling of the genomic DNA would also play a critical role. Recently Geisel and Gerland [9] investigated the role of strength of protein-protein interactions among TFs on transcription regulation using a two TFs combinatorial model system. In their model, combination of two different TFs viz. $TF_1$ and $TF_2$ regulate the same gene in an AND-logic mode. This study suggested that both weak and strong level of protein-protein interactions among these TFs are more efficient in regulating the transcription initiation rather than the intermediate level of protein-protein interactions [9]. The cooperativity among these two TFs can be measured by the dimensionless number $w = \exp(-\mu)$ where $\mu$ in the overall free energy (measured in the units of $k_BT$) associated with the protein-protein interactions among the combinatorial TFs in the system [9].

Here the weakest level of protein-protein interactions among TFs represents a monomer pathway of protein-DNA interactions and a strongest level of protein-protein interactions among TFs represents a heterodimer pathway of site-specific protein-DNA interactions (**Fig. 1A** and **B**). In the monomer pathway, the protein-protein interactions are very weak so that TFs ($TF_1$ and $TF_2$) search for their cognate sites on DNA in an independent manner. In the dimer pathway, the protein-protein interactions are very strong so that the $TF_1$-$TF_2$ heterodimer as a whole searches for the sequentially located cognate sites on DNA and strongly binds there.

Clearly there is no configuration hindrance in both monomer and dimer pathways since TFs will be homogenously populated either as monomers or heterodimers in these extreme cases. However in the intermediate level of protein-protein interactions, there will be significant fractions of population associated with both monomers and heterodimers and there is a dynamic equilibrium between these subpopulations. This means that there is a possibility for the arrival of a $TF_1$-$TF_2$ dimer towards the cognate sites while $TF_1$/$TF_2$ monomer is present at its specific binding site. Such steric hindrances would increase the number of trials required for the complete assembly of the combinatorial TFs at CRMs [9].

Apart from these monomer and dimer pathways, there is also a possibility for the recruitment pathway [1, 10-12] in which the TF which arrived at its cognate site earlier will recruit the TF corresponding to its adjacent binding site via protein-protein interactions (**Fig. 1C**). Clearly recruitment pathway is a special form of the monomer pathway. The random searching mode is same in both these mechanisms however they differ in the way TFs interacts with their target sites upon arrival. Typical example for the recruitment pathway is the p53 mediated transcription regulation [13]. Here binding of p53 tetramer with its response element will in turn recruit diverse transcription coregulators such as histone modifying enzymes, chromatin remodeling factors, subunits of the mediator complex, and components of general transcription machinery and preinitiation complex (**PIC**) to modulate RNAPII activity at target loci [14].

In the absence of recruitment, there are chances for the monomer pathway system to get trapped in a wrong configuration which in turn strictly requires hopping or dissociation type dynamics of TFs. For example consider the CRM of a gene i.e. $X_1X_2X_3X_4$ for four TFs viz. $TF_1$, $TF_2$, $TF_3$ and $TF_4$ respectively and assume that $TF_1$ has already bound at $X_1$. In recruitment pathway $TF_1$ will in turn recruit $TF_2$ towards $X_2$ and then $TF_2$ will recruit $TF_3$ towards $X_3$ and so on. In the monomer pathway since TFs bind independent of each other there is a chance of getting a configuration $X_1$-$TF_1$: $X_2$: $X_3$-$TF_3$: $X_4$. This configuration requires crossing of $TF_2$ over $TF_1$ or $TF_3$ to reach $X_2$ [7].

Experimental studies suggested that the recruitment pathway of regulation is the universal one especially in case of eukaryotic systems [10]. In the present context, monomer to intermediate level of protein-protein interactions can facilitate the recruitment pathway especially when the number of combinatorial TFs is much higher than two. Under such conditions it is also meaningless to consider a hetero-oligomer pathway made out of several TFs since the probability associated with such many body collisions is negligible [15]. Though the monomer and dimer pathways are more efficient than the intermediate regime for the two TFs system it is still not clear whether these pathways are efficient for the combinatorial regulation with higher number of TFs. It is also still not clear about the physical basis of the recruitment pathway and the origin of cooperative effects. In this paper using a combination of theoretical and simulation tools we will address these issues in detail.

## 2. Theory

Let us consider a linear DNA lattice of size $N$ bps which contains the CRM for $TF_1$. Upon following the two step 3D1D diffusion model [16-19] of site-specific DNA-protein interactions, the overall search time or mean first passage time (MFPT) $\tau_{S,U}$ associated with $TF_1$ to find its CRM located on DNA via a combination of 3D and 1D diffusion can be written as follows [20].

$$\tau_{S,U} \simeq \left[[P_{BTF1}]\left(k_{fa} + k_{fX}/(1 + k_r \lambda \eta_U)\right)\right]^{-1}; \; \lambda = N/U; \; \eta_U = U^2/12d_o \quad [1]$$

In this equation $P_{BTF1}$ (M, mols/lit) is the concentration of $TF_1$ in cytoplasm, $k_{fa}$ (M$^{-1}$ s$^{-1}$) is the bimolecular rate constant associated with the direct site-specific binding of $TF_1$ via 3D diffusion, $k_{fX}$ is the overall non-specific binding rate and $k_r$ (s$^{-1}$) is the dissociation rate of nonspecifically bound $TF_1$. Further $\lambda$ is the number of association-scan-dissociation cycles required by $TF_1$ to scan the entire DNA and $\eta_U$ is the overall average time that is required by $TF_1$ to scan $U$ bps of DNA via 1D diffusion before dissociation. Here $U$ is a random variable that will take different values in each association-scan-dissociation cycle. The probability density function associated with the 1D diffusion lengths $U$ can be written as follows [20].

$$p_U(U) \simeq 2U e^{-(U/\Phi_A)^2}/\Phi_A^2; \; \Phi_A = \sqrt{12d_o/k_r} \quad [2]$$

Here $\Phi_A$ is the maximum possible 1D diffusion length associated with the nonspecifically bound $TF_1$ on DNA that is measured in bps where 1bps ~ 3.4 x 10$^{-10}$ m and $d_o$ (bps$^2$s$^{-1}$) is the 1D diffusion coefficient associated with the dynamics of $TF_1$. Understanding the way by which **Eq. 1** is modified depending on the number of TFs involved in a combinatorial regulation and the underlying binding modes viz. monomer, hetero-oligomer and recruitment pathways are the central topics of the following sections.

When the TFs of interest move with a hop size of $k$ then we find $d_o = l_d^2 \sum_{i=-k}^{k} p_i w_i i^2$ where $i \in Z$, $w_{\pm i}$ are the microscopic transition rates associated with the forward and reverse movements of TFs on DNA and $p_{\pm i}$ are the corresponding microscopic transition probabilities [18, 21]. Here the step length $il_d$ is measured in terms of bps. We have defined $l_d$ = 1bps. Since the dynamics at the DNA-protein interface involves segmental motion of DNA binding domains of TFs one can assume protein folding rate limit [22] for the transition rates as $w_{\pm 1}$ ~ 10$^6$ s$^{-1}$. Noting that $p_{\pm 1}$ ~ ½ for an unbiased 1D random walk, one finds that $d_o$ ~ 10$^6$ bps$^2$s$^{-1}$ for a sliding type dynamics of TFs for which $k$ = 1. Approximately this is the experimental value of 1D diffusion coefficient associated with the sliding dynamics of TFs on DNA [20, 23, 24]. For an arbitrary hop size of $k$, when $p_{\pm i} = 1/2k$ and $\langle w_{\pm i} \rangle = \phi$ then one finds that $d_o = l_d^2 \phi (k+1)(2k+1)/6$.

## 2.1. Random walks with random hop size

Let us consider the DNA chain as a linear lattice confined within $(x_L, x_R)$ (**Fig. 2**) where $|x_R-x_L| = N$. Inside this lattice we consider a single TF1 that is searching for its CRM located at $x = x_A$. Here $(x_L, x_R)$ are reflecting boundaries and $x_A$ is the absorbing boundary. The Langevin type stochastic differential equation that describes the dynamics of such TF molecule can be written as follows [8, 21, 25-27].

$$dx/dt = \sqrt{d_o}\Gamma_t;\ t = t_0;\ x = x_Z;\ \langle\Gamma_t\rangle = 0;\ \langle\Gamma_t\Gamma_{t'}\rangle = \delta(t-t') \qquad [3]$$

In this equation $x$ is the position of TF1 at time $t$ and $\Gamma_t$ is the Gaussian white noise whose mean and covariance properties are defined as in **Eq. 3**. The probability density function associated with the dynamics of TF1 on a linear lattice of DNA obeys the forward Fokker-Planck equation (FPE) which can be written along with the boundary conditions as follows.

$$\partial_t P(x,t|y_Z,t_0) = (d_o/2)\partial_x^2 P(x,t|y_Z,t_0);\ [\partial_x P(x,t|y_Z,t_0)]_{x=x_L,x_R} = 0;\ P(x_A,t|y_Z,t_0) = 0 \qquad [4]$$

Here $P(x,t|y_Z,t_0)$ is the probability of observing TF1 at position $x$ at time $t$ with the condition that it was at $x = y_Z$ at $t = t_0$. Apart from the boundary conditions given in **Eq. 4** one also needs to set the initial condition as $P(x,t_0|y_Z,t_0) = \delta(x-y_Z)$. To simplify our analysis and other computations we use the following scaling transformations so that the dynamical variables in **Eq. 4** become dimensionless [8].

$$\langle w_{\pm i}\rangle = \phi;\ \tau = \phi t;\ X = x/l_d;\ Y_Z = y_Z/l_d;\ D_o = d_o/\phi l_d^2 = (k+1)(2k+1)/6 \qquad [5]$$

When $k = 1$ then $D_o = 1$. Upon rescaling the variables in **Eq. 4** as in **Eq. 5** we obtain the following Fokker-Planck equation in dimensionless form.

$$\partial_\tau P(X,\tau|Y_Z,\tau_0) = (D_o/2)\partial_X^2 P(X,\tau|Y_Z,\tau_0) \qquad [6]$$

The corresponding initial and boundary conditions are as follows.

$$P(X,\tau_0|Y_Z,\tau_0) = \delta(X-Y_Z);\ [\partial_X P(X,\tau|Y_Z,\tau_0)]_{X=X_L,X_R} = 0;\ P(X_A,\tau|Y_Z,\tau_0) = 0 \qquad [7]$$

Here $X_Q = x_Q/l_d$ where the subscript $Q$ can be $Q = L, R, A$. Let us first consider a situation where $k = 1$ and $X_L < Y_Z < X_A < X_R$. Clearly the probability of finding TF1 in $(X_A, X_R)$ is zero. Under such conditions the mean first passage time (MFPT) associated with the escape of the 1D random walker from the interval $(X_L, X_A)$ through the absorbing point $X_A$ starting from $Y_Z$ will obey the following backward type FPE with appropriate boundary conditions.

$$D_o d_X^2 \Pi_S = -2;\ [d_X \Pi_S]_{X=X_L,X_R} = 0;\ [\Pi_S]_{X=X_A} = 0;\ \Pi_S = (X_A^2 - Y_Z^2 - 2X_L(X_A - Y_Z))/D_o \qquad [8]$$

Mechanisms of combinatorial binding of transcription factors

Similarly when $k = 1$ and $X_L < X_A < Y_Z < X_R$ then the probability of finding $TF_1$ in $(X_L, X_A)$ will be zero and one finds that $\Pi_S = \left(X_A^2 - Y_Z^2 - 2X_R(X_A - Y_Z)\right)/D_o$. Here MFPT will be measured in terms of dimensionless number of simulation steps. The results presented in **Eqs. 3-8** are standard and well known [19, 21, 28, 29]. Now we introduce $n$ number of TFs into the linear lattice (e.g. **Fig. 2** where $n = 4$). We denote them as $TF_1$, $TF_2$....$TF_n$. The corresponding CRMs are $X_A = X_1X_2....X_n$ which are all confined inside the lattice $(X_L, X_R)$. The complete assembly of $n$ TFs over these CRMs starting from $Y_Z = Y_1, Y_2...Y_n$ as $X_1$-$TF_1$: $X_2$-$TF_2$: ....: $X_n$-$TF_n$ will subsequently lead to the initiation of transcription.

When TFs perform a pure 1D sliding type dynamics with $k = 1$, then a presorted initial condition is mandatory [7] for the correct assembly of TFs over CRMs. That is to say, we must arrange these $n$ TFs in the order as $TF_1$, $TF_2$...$TF_n$ along the DNA lattice. Particularly when $X_A = X_1X_2...X_n$, then the required presorted initial condition will be $Y_Z = Y_1Y_2...Y_n$ irrespective of whether $X_L < Y_Z < X_A < X_R$ or $X_L < X_A < Y_Z < X_R$. Although the exact initial positions $Y_Z = Y_1, Y_2...Y_n$ of TFs on the DNA lattice can be arbitrary, the inequality condition $Y_1 < Y_2 < ... < Y_n$ is mandatory for the successful assembly of TFs over CRMs which otherwise warrants a physical dissociation-association or hopping type dynamics of TFs. To compute the MFPT associated with the complete assembly of $n$ TFs at $X_A$ starting from $Y_Z$ one needs to consider the boundary conditions pertained to the dynamic reflections among TFs which are mandatory to enforce the excluded volume effects. The backward FPE in **Eq. 8** cannot be solved analytically with several such boundary conditions though one can stochastically simulate such systems of $n$ TFs.

When $X_L < Y_Z < X_A < X_R$ and $\Delta = |X_A - Y_Z|$ is large, then earlier stochastic random walk simulation studies [7] showed a scaling as $\Pi_{n,1} \simeq \Pi_{1,1} n^\alpha$ where $\Pi_{1,1} = \left(X_1^2 - Y_1^2\right)$ for $X_L = 0$. Here $\Pi_{n,1}$ is the MFPT associated with complete assembly of $n$ TFs at their respective CRMs via pure sliding mode of dynamics, $\Pi_{1,1}$ is the MFPT associated with a single TF to find its target site on DNA via sliding mode of dynamics. Linear least square fitting procedures suggested that $\alpha \sim 2/5$. In the notation for MFPT as $\Pi_{n,k}$, the subscript $n$ denotes the number of TFs in the system and the subscript $k$ denotes the hop size associated with the dynamics of TFs. For a sliding type dynamics we have $k = 1$.

### 2.2. Monomer pathway of combinatorial regulation
In the monomer pathway all the $n$ TFs independently search for their cognate sites. Additionally binding of $n$ TFs at their CRMs will be an incoherent event. Let us first consider a situation where $X_L < Y_Z < X_A < X_R$, $X_L = 0$ and $k = 1$. We assume that initial positions as well as target sites of $n$ TFs are sequentially located on DNA lattice. When the initial positions of $n$ TFs are close to the respective CRMs, then the exponent $\alpha$ seems to be strongly dependent on $n$ as well as the average distance of CRMs from the initial positions of TFs. Detailed stochastic random walk simulations and subsequent nonlinear least square fittings (**Figs. 3A, C** and **E**) suggested the following relationship for the MFPT associated with the complete assembly of $n$ TFs.

$$\Pi_{n,1} \simeq \Pi_{1,1} n^{\alpha_{n,\Delta,1}}; \ \Pi_{1,1} = \left(X_1^2 - Y_1^2\right); \ \alpha_{n,\Delta,1} \simeq \gamma + \beta \ln n; \ \gamma = a + b\exp(-c\Delta); \ \Delta = |X_1 - Y_1| \qquad [9]$$

Here the expression for $\Pi_{1,1}$ is derived from **Eq. 8** by substituting $X_L = 0$ where $D_o = 1$ for $k = 1$ as in **Eq. 5**. The exponent $\alpha_{n,\Delta,k}$ is strongly dependent on $n$, $\Delta$ and hop size $k$ where $k = 1$ in the present context and $\Delta$ is the absolute average distance between the initial positions of TFs on DNA and their CRMs. One can rewrite **Eq. 9** in a linearized form as follows.

$$y \simeq y_0 + \gamma z + \beta z^2; \ y = \ln \Pi_{n,1}; \ y_0 = \ln \Pi_{1,1}; \ z = \ln n \quad [10]$$

Now we consider a situation where $X_L < Y_Z < X_A < X_R$, $X_L = 0$ and $k > 1$. When $k > 1$, then the integral solution in **Eq. 8** for $n = 1$ is no more valid and the subdomains $(X_L, X_A)$ and $(X_A, X_R)$ will be dynamically connected. That is to say, even though $X_L < Y_Z < X_A < X_R$, TFs can hop over from $(X_L, X_A)$ to $(X_A, X_R)$ without hitting the absorbing point $X_A$ when $k > 1$. The probability of such hopping events will be directly proportional to $p_k \simeq (1 - 1/k)$. When the hop size is such that $k > n$, then the presorted initial conditions of TFs is not necessary and subsequently **Eq. 9** will be modified as follows.

$$\Pi_{n,k} \simeq \Pi_{1,k} n^{\alpha_{n,\Delta,k}}; \ \Pi_{1,k} = \Pi_{1,1}/D_o + X_R(1 - 1/k); \ \alpha_{n,\Delta,k} \simeq \gamma + \delta/(\ln k - \ln n) + \beta \ln n \quad [11]$$

Here $D_o = (k+1)(2k+1)/6$ and the exponent $\gamma$ depends on $\Delta$ in an exponential manner as in **Eq. 9**. Similar to **Eq. 10** one can linearize **Eq. 11** as follows.

$$y \simeq y_0 + \left[\gamma + \delta/(\ln k - z)\right]z + \beta z^2; \ y = \ln \Pi_{n,k}; \ y_0 = \ln \Pi_{1,k}; \ z = \ln n \quad [12]$$

Nonlinear least square fitting analysis (**Figs. 3B**, **D** and **F**) of the stochastic simulation data with **Eqs. 9** and **11** suggests that the coefficient $b$ in the definition of $\gamma$ decreases with the hop size $k$ as the way the coefficient $\delta$ behaves in **Eq. 11**. This observation suggested the following final linearized form.

$$y \simeq y_0 + \left[a + (b\exp(-c\Delta) + \delta)/(\ln k - z)\right]z + \beta z^2; \ y = \ln \Pi_{n,k}; \ y_0 = \ln \Pi_{1,k}; \ z = \ln n \quad [13]$$

Since $\beta \simeq 0$, for sufficiently large values of $\Delta$ and $k$ **Eq. 13** reduces to $y \simeq y_0 + az$ as observed in the earlier studies [7]. Because of the incoherent binding of TFs at their CRMs $\eta_U$ in **Eq. 1** transforms as $\eta_U \to n^a \eta_U$. This means that at large values of $n$, the 3D1D mode of searching of TFs for their CRMs associated with the monomer pathway will asymptotically reduce to a pure 3D-only mode of searching. Here one should note both 3D-only and 3D1D diffusion modes of target search operate in parallel [20]. This means that the monomer pathway will not be an efficient one for large values of $n$ in the combinatorial regulation of TFs. Upon considering the scaling up property of the 1D search time $\eta_U$ with respect to the number of TFs $n$ one finds that $\Phi_A \to \Phi_A n^{-a/2}$. As a result of the dynamic reflections and spatial confinement of TFs [30], the maximum possible 1D sliding length associated with the dynamics of TFs on DNA decreases in a power law manner as the number of TFs in a combinatorial regulation increases.

Here the hop size $k$ is positively correlated with the degree of condensation or supercoiling of the DNA polymer. From **Eq. 5** one finds that the 1D diffusion coefficient $D_o$ is directly proportional to square of $k$. One should note that the effective $D_o$ can be enhanced by increasing the degree of condensation (hence by increasing $k$) of the DNA polymer only up to certain extent since $D_o \leq D_t$ where $D_t$ is the 3D diffusion coefficient associated with the translational dynamics of TFs. Here one should note that the probability density function associated with the distribution of $k$ will be $p_{\pm i} = 1/2k$ only for small values of $k$. The probability associated with the occurrences of large values of $k$ will be very less. At those values of $k$ for which $D_o \sim D_t$, 1D and 3D diffusion dynamics of TFs will be physically indistinguishable.

### 2.3. Hetero-oligomer pathway of combinatorial regulation

In the hetero-oligomer pathway, there will be a formation of protein-protein complex of $n$ TFs. Subsequently this hetero-oligomer complex as a whole will search for the sequentially located CRMs on DNA via 3D1D mode of searching (**Fig. 1B**). The site-specific binding rate via 3D-only diffusion mode of searching ($k_{fa} \simeq k_t \chi/8$ [20] in **Eq. 1**) will be approximately independent on the number of TFs in the protein-protein complex. Here $k_t$ is the Smolochowski bimolecular collision rate limit and $\chi$ is the dimensionless multiplication factor corresponding to the overall electrostatic interactions at the DNA-protein interface. However the nonspecific collision rate $k_{fX} \simeq k_t N/R_P$ [20] in **Eq. 1** for a relaxed conformational state of DNA will be strongly dependent on $n$ since the radius of gyration $R_P$ of the protein-protein complex of $n$ TFs is strongly dependent on $n$. In an extreme situation one finds that $R_P \propto n$.

In the same way, the overall 1D diffusion coefficient $d_o$ associated with the protein-protein complex of $n$ TFs will be rescaled as $d_o \to d_o/n$ which means that the rescaling as $\eta_U \to n\eta_U$ will be true for a hetero-oligomer pathway of combinatorial regulation. In this analysis we have not considered the improbable nature of the $n$-body collisions when $n$ is very large. Upon considering all these facts, one can conclude that the overall type of searching of TFs for their cognate CRMs will be driven from 3D1D to 3D-only mode much faster in hetero-oligomer pathway (as $n^{-2}$) than the monomer pathway (as $n^{-a}$ where $a < 1$).

### 2.4. Recruitment pathway of combinatorial regulation

Now we consider the recruitment pathway. This is a special form of the monomer pathway. Recruitment mode differs from the monomer mode in the way by which TFs interacts with their target sites upon arrival. In the recruitment pathway, the TF which arrives first will recruit the adjacent TFs in a sequential manner either with or without cooperative effects. In the absence of cooperativity, the free energy released from the protein-protein interactions among TFs does not propagate towards stabilizing the DNA-protein interactions of the incoming TFs. In the presence of cooperative effects, the extent of stabilization of the complex with $m+1$ number of TFs will be directly proportional to $m$. To model the recruitment process, we assume that all the $n$ TFs (including the TF which arrives foremost) perform a simultaneous random search as in the monomer pathway and interact with their respective CRMs in coordinated manner via two step 3D1D mode with an average search time of $\tau_{S,U}$ as in **Eq. 1**.

Mechanisms of combinatorial binding of transcription factors

### 2.4.1. Passive recruitment without cooperative effects

The passive sequential assembly of $n$ TFs over their respective CRMs can be described by the following birth-death master equation.

$$\partial_t P(u,t|u_0,t_0) = k_+ P(u-1,t|u_0,t_0) + k_- P(u+1,t|u_0,t_0) - (k_+ + k_-)P(u,t|u_0,t_0) \qquad [14]$$

Here $u$ is the number of TFs out of $n$ assembled over CRMs at time $t$, $k_+$ and $k_-$ are the respective forward and reverse rate constants and $P(u, t |u_0, t_0)$ is the probability of observing $u$ number of TFs assembled at time $t$ with the condition that there were $u_0$ number of TFs assembled at time $t = t_0$. Clearly $k_+ \simeq 1/\tau_{S,U}$ of **Eq. 1**. The initial and boundary conditions can be given as follows.

$$P(u,t_0|u_0,t_0) = \delta(u-u_0); \; k_- P(0,t|u_0,t_0) = k_+ P(1,t|u_0,t_0); \; P(n,t|u_0,t_0) = 0 \qquad [15]$$

Using the backward master equation corresponding to **Eq. 14** one can derive [21, 25-27] the MFPT associated with the complete passive assembly of $n$ TFs over their CRMs starting from $u_0 = 1$ in a sequential manner as follows.

$$\tau_S = \sum_{u=1}^{n} \phi(u) \sum_{w=1}^{u} (k_+ \phi(w))^{-1} = \left(K_A^{n+1} - K_A(n+1) + n\right) \Big/ k_+ (1-K_A)^2 \; ; \; \phi(u) = \prod_{w=2}^{u} K_A \qquad [16]$$

Here $u = 1$ is a reflecting boundary and $u = n$ is the absorbing boundary and $K_A = (k_-/k_+)$. From **Eq. 16** one finds that $\lim_{K_A \to 1} \tau_S \simeq n(1+n)/2k_+$ and $\lim_{K_A \to 0} \tau_S \simeq n/k_+$. These results suggest that in case of passive recruitment without cooperative effects, the overall search time required for the complete assembly of $n$ TFs at their respective CRMs will be scaled up with $n$ as $\tau_{S,U} \to n\tau_{S,U}$ or $\tau_{S,U} \to n^2 \tau_{S,U}$ depending on the condition that $K_A = 0$ or $K_A = 1$ respectively. Here $K_A = 1$ represents a pure diffusion like assembly of TFs and $K_A = 0$ represents a directed-walk like assembly or passive recruitment pathway. These results suggest that the recruitment pathway without cooperative effects is no more efficient than the monomer or hetero-oligomer pathways.

### 2.4.2. Active recruitment with cooperative effects

In the presence of cooperative effects, the master equation **Eq. 14** will be modified as follows.

$$\partial_t P(u,t) = k_+(u-1)P(u-1,t) + k_-(n-u-1)P(u+1,t) - (k_+ u + k_-(n-u))P(u,t) \qquad [17]$$

In this equation the overall probability associated with the transition $u \to u+1$ will be directly proportional to $u$. In the same way the transition probability associated with $u \to u-1$ will be directly proportional to $n-u$ i.e. number of free SBSs. Upon solving the backward master equation corresponding to **Eq. 17** with boundary conditions given in **Eq. 15** one obtains the following expression for the overall MFPT.

$$\tau_S = \sum_{u=1}^{n} K_A^u \left( \xi \,_2F_1([1,1],[2-n],-K_A^{-1}) + \phi \,_2F_1([1,u+1],[u+2-n],-K_A^{-1}) \right) \qquad [18]$$

In this equation $\tau_S$ is the MFPT associated with the complete assembly of $n$ TFs at their CRMs in a sequential manner in the presence of cooperative effects mediated through protein-protein interactions among TFs. Here $_2F_1$ is the hypergeometric function [31, 32] and various parameters are defined as follows.

$$\xi = (-1)^u \Gamma(u+1-n)/k_-\Gamma(u+1)\Gamma(1-n)(n-1); \; \phi = \Gamma(u+1-n)/K_A^{u+1}k_+\Gamma(u+2-n)$$

The hypergeometric function of the type $_2F_1$ is defined as follows.

$$_2F_1([a,b],g,w) = \sum_{m=0}^{\infty} w^m (a)_m (b)_m / m!(g)_m; \; (h)_q = \Gamma(h+q)/\Gamma(h)$$

The complicated expression given by **Eq. 18** can be simplified by the Fokker-Planck equation (FPE) formalism. The forward FPE corresponding to **Eq. 17** can be written as follows.

$$\partial_t P(u,t|u_0,t_0) = -\partial_u (A(u)P(u,t|u_0,t_0)) + \partial_u^2 (B(u)P(u,t|u_0,t_0))/2 \quad [19]$$

Here the drift and diffusion coefficients can be written as follows.
$$A(u) = k_+ u - k_-(n-u); \; B(u) = k_+ u + k_-(n-u)$$

Using the backward type FPE corresponding to **Eq. 19** one can obtain the mean first passage time associated complete assembly of $n$ TFs over their CRMs as follows.

$$\tau_S \simeq (2/k_+) \int_1^n (H(y)/\Phi(y)) dy; \; H(y) = \int_1^y (\Phi(w)/(w+K_A(n-w))) dw \quad [20]$$

In this equation various functions and parameters are defined as follows.

$$\Phi(q) = \exp\left(2\int_1^q p(w) dw\right); \; p(w) = A(w)/B(w) = (w - K_A(n-w))/(w + K_A(n-w))$$

**Eq. 20** suggests that as $K_A$ tends towards zero, the overall time that is required for the complete assembly of $n$ TFs approximately scales with $n$ in a logarithmic way that can be demonstrated as follows. Upon defining $\lim_{K_A \to 0} \tau_S = \tilde{\tau}_S$ one can derive the following expression [32].

$$\tilde{\tau}_S = \left(\ln n + e^{-2n}(\text{Ei}(2) - \text{Ei}(2n))\right)/k_+; \; \text{Ei}(h) = \int_{-\infty}^h e^x x^{-1} dx \quad [21]$$

This equation suggests that for sufficiently large values of $n$, in the presence of cooperative effects the change in the overall time that is required for the complete assembly of $n$ TFs with respect to $n$ will be independent of $n$ since $\lim_{n \to \infty} \partial_n \tilde{\tau}_S = 0$. From **Eq. 21** one finds that at sufficiently large values of $n$ the overall search time associated with the binding of $n$ TFs will be scaled up with $n$ as $\tau_{S,U} \to \tau_{S,U} \ln n$ in the presence of cooperative effects. These results suggest that the recruitment pathway with cooperative effects is the most efficient one with respect to the TFs mediated combinatorial regulation of transcription.

Mechanisms of combinatorial binding of transcription factors

## 3. Stochastic random walk simulations

To understand the site-specific binding of multiple TFs at their CRMs, we performed mean first passage time calculations over detailed random walk simulations. We considered a linear lattice of size $N = 151$ bps in which there are $n$ number of random walkers (TFs) where $n$ was iterated from 1 to 25. Here both the left boundary at $X_L = 0$ and right boundary at $X_R$ are a reflecting ones and $|X_R - X_L| = N$. Depending on $n$, the initial positions of random walkers were set from $Y_Z$ to $Y_Z + (n-1)$ where $Y_Z = (Y_1, Y_2 \ldots Y_n)$ and $Y_1$ was iterated from 1, 50, 75, 90 and 95. The corresponding sequentially located CRMs of TFs $X_A = (X_1, X_2 \ldots X_n)$ were set from 105 to 130. Both $Y_Z$ and $X_A$ are confined inside $(X_L, X_R)$.

We computed the MFPT associated with the complete assembly of $n$ TFs at $X_A$ starting from $Y_Z$. All the simulations were carried out over dimensionless space as defined in **Eq. 5**. Here MFPT will be measured in terms of number of simulation steps. When the hop size $k = 1$, then a TF of interest will be reflected back upon encountering another TF along its trajectory on the linear lattice. Any two TFs cannot occupy the same lattice position on DNA to enforce the excluded volume effect. When $k = 2$, then a TF can hop over another TF in a row. When $k = 3$, then a TF of interest can hop over other two TFs in a row and so on.

Cleary a presorted initial condition $Y_1 < Y_2 < \ldots < Y_n$ is mandatory when $k < n$ which otherwise leads to trapping of the system in a wrong configuration. This will drive the MFPT towards infinity. To explain this we consider $n = 2$ and $k = 1$. When $X_A = X_1 X_2$, then this requires a presorted initial condition as $X_L < Y_1 < Y_2 < X_A < X_R$ or $X_L < X_A < Y_1 < Y_2 < X_R$ for the successful assembly of both TFs at their respective CRMs. Suppose let us assume that $X_L < Y_2 < Y_1 < X_1 < X_2 < X_R$. Under pure sliding mode of dynamics with $k = 1$, $TF_1$ can bind with $X_1$ and $TF_2$ needs to hop over $TF_1$ to reach $X_2$ which in turn requires $k > 1$. When $n = 1$ and $k = 1$, then the MFPT associated with the binding of $TF_1$ at $X_1$ starting from $Y_1 = 1$ will be $\Pi_{1,1} = (X_1^2 - Y_1^2)$ = $105^2 - 1^2 = 11024$ and $\ln \Pi_{1,1} = 9.3078$. The y-intercept of black solid line in **Fig. 3A** agrees well with this prediction.

When $n$ is increased, then detailed nonlinear least square fitting procedures using Marquardt-Levenberg algorithm [33, 34] revealed that $\Pi_{n,1} \simeq \Pi_{1,1} n^{\alpha_{n,\Delta,1}}$ where $\alpha_{n,\Delta,1} = \gamma + \beta \ln n$. In this fit function both $\Pi_{1,1}$ and $\gamma$ are strongly dependent on the distance $\Delta = |X_1 - Y_1|$. One finds from **Eq. 8** that $\Pi_{1,1} = X_1^2 - (X_1 - \Delta)^2$. Results presented in **Fig. 3C** agree well with this prediction. Further analysis in **Fig. 3E** suggested that the exponent $\gamma$ depends on $\Delta$ in an exponential manner as $\gamma = a + b \exp(-c\Delta)$ where the fitted parameters were $a = 0.46 \pm 0.05$, $b = 0.72 \pm 0.06$ and $c = 0.03 \pm 0.001$ at a confidence level of 0.95. These results suggest $\alpha_{n,\Delta,1} \simeq a + b \exp(-c\Delta) + \beta \ln n$. Since $\beta$ is close to zero, one finds that $\Pi_{n,1} \simeq \Pi_{1,1} n^a$ at sufficiently large values of $\Delta$ for $k = 1$.

Now we fixed $n$ and increased the hop size $k$. Clearly when $k < n$, then the presorted initial condition is mandatory which otherwise drives the MFPT towards infinity. When $k > n$, then the presorted initial condition is not mandatory. In these particular simulations we set $k = 25$ and iterated $n$ from 1 to 24. For $n = 1$ and $k = 25$, one finds that $\Pi_{1,k} \simeq \Pi_{1,1}/D_o + X_R(1-1/k)$ where explicitly one finds $D_o = 212.5$ from **Eq. 5** for the hop size $k = 25$. Upon substituting the value of $X_R = 151$ we obtain $\ln \Pi_{1,25} \simeq 5.2722$ and clearly $\ln \Pi_{1,\infty} \simeq \ln X_R = 5.0173$. The $y$-intercept of black solid line that corresponds to $X_1 = 105$ and $Y_1 = 1$ in **Fig. 3B** agrees well with this result.

As we increase $n$ from 1 to 25, using detailed nonlinear least square fitting procedures one finds that $\Pi_{n,k} \simeq \Pi_{1,k} n^{\alpha_{n,\Delta,k}}$ where $\alpha_{n,\Delta,k} = \gamma + \delta/\ln(k/n) + \beta \ln n$ and the exponent $\gamma$ strongly depends on $\Delta$ in an exponential manner as $\gamma = a + b\exp(-c\Delta)$. The nonlinear lest square fitted parameters were $a = 0.45\pm0.1$, $b = 0.15\pm0.1$ and $c = 0.02\pm0.004$ at a confidence level of 0.95. Clearly $b$ in the expression for $\gamma$ approximately decreases as $b \to b/\ln(k/n)$ when the hop size $k > n$ increases. This suggested the final expression for MFPT as $\Pi_{n,k} \simeq \Pi_{1,k} n^{\alpha_{n,\Delta,k}}$ where we found the expression for the exponent as $\alpha_{n,\Delta,k} = a + (b\exp(-c\Delta) + \delta)/\ln(k/n) + \beta \ln n$. This expression predicts the MFPT well for $k > n$. For sufficiently large values of $k$ and $\Delta$ and noting that $\beta$ and $\delta$ are close to zero, one finds that $\Pi_{n,\infty} \simeq X_R n^a = 642.76$ with respect to the current simulation settings where $n = 25$, $X_R = 151$ and $a = 0.45$. Simulation data presented in **Figs. 4A-B** agree well with this prediction.

## 4. Results and Discussion

TFs play critical roles in regulating various genes in prokaryotes and eukaryotes. Most of the eukaryotic TFs regulate the transcription of various genes in a combinatorial manner which is essential for handling the speed-fidelity issues related to the large genome sizes and nuclear volumes. Binding of combinatorial TFs at the corresponding CRMs (enhancers) results in the formation of ETF complex which in turn bends of the DNA polymer [3, 10]. This results in the distal action of ETF complex over the RNAPII-promoter complex either via looping of DNA or tracking of ETF along DNA [5, 6, 35]. Site specific binding of multiple TFs at sequentially located CRMs is a typical example for the random search under crowded environments.

There are at least three different mechanisms by which the combinatorial TFs can locate their respective CRMs viz. monomer, hetero-oligomer and recruitment pathways (**Fig. 1**). In the monomer pathway, the protein-protein interactions among the combinatorial TFs are negligible and they locate their respective CRMs in an independent manner. However in the process of random searching, the dynamics of a TF of interest will be strongly influenced by various molecular crowding effects viz. dynamic reflections and spatial confinements due roadblocks which in turn result in an anomalous type diffusion of TFs [30]. Here one should note that the conformational state of DNA can modulate the extent of crowding effects. Condensed conformational state of DNA helps a TF of interest to circumvent other roadblock TFs without dissociation from DNA.

Our detailed simulation results on the monomer pathway suggested that (**I**) the crowding effects are negatively correlated with the average initial distance of TFs ($\Delta$) from their CRMs (**II**) the overall MFPT associated with the binding of $n$ TFs with their CRMs will scale with $n$ in a power law manner (**III**) the power law exponent depends on the hop size ($k$) associated with dynamics of TFs on DNA, initial distance of TFs from their CRMs and $n$ (**IV**) minimum degree of supercoiling or condensation of DNA is mandatory to ensure the condition that $k > n$ and (**V**) when the hop size $k$ is close to $n$, then the power law exponent increases steeply that is evident from **Eq. 11**. Observation **I** is evident from that fact that the power law exponent associated with the increase in MFPT increases as the distance between the initial positions of TFs from their CRMs decreases (**Figs. 3C** and **E**). Observation **V** is a reasonable one since the probability associated with the escape of the system from wrong configurations decreases steeply as the hop size $k$ approaches $n$.

For sufficiently large values of $\Delta$ and $k$ our results (along with the expression for the search time predicted by 3D1D model as in **Eq. 1**) suggested that as $n$ increases the overall search time associated with the binding of $n$ TFs with their CRMs increases as $n^a$ (**Figs. 4A** and **B**). Here the exponent is nonnegative and $a < 1$. These mean that the time required for the site specific binding of $n$ TFs will be $n^a$ times higher than the time that is required for the binding of single TF with its CRM. This is eventually the cost of combinatorial regulation with several TFs through the monomer pathway.

Although the hetero-oligomer pathway has been shown to be a feasible one for $n = 2$ [9], it is not an efficient one for the combinatorial regulation with several TFs. This is mainly because such $n$-body collisions are improbable and the overall search time associated with the binding of $n$ TFs with their CRMs increases with $n$ as $n^2$. Recruitment pathway differs from monomer pathway only in the way by which combinatorial TFs interact with their target binding sites. Recruitment pathway of combinatorial regulation of TFs seems to be a universal one in eukaryotic systems [1, 10]. In this mode, the TF which arrived at its CRM first will recruit the TFs corresponding to its adjacent CRMs and so on. In such process of recruitment, the protein-protein interactions among adjacent TFs may or may not propagate as cooperative effects. Our detailed analysis suggested that the recruitment pathway with propagating cooperative effect is the most efficient one since the overall search time associated with the binding of $n$ TFs with their CRMs under such scheme scales with $n$ in a logarithmic manner. In this coordinated scheme the cost of combinatorial binding of TFs will be partially compensated by the propagation of protein-protein interactions among TFs as cooperative effect.

The coordinated sequential binding of TFs in the recruitment pathway can be implemented by a change in the conformational state of DNA upon binding of the first TF. For example binding of the first TF can bend the DNA polymer so that it facilitates only the binding of adjacent TFs and destabilizes the random binding of other distal TFs. Clearly such mechanisms are possible only when the first TF molecule is large and its site specific binding energy is high enough to bend the DNA polymer. Typical example for such recruitment system is the p53 mediated transcription regulation. Otherwise a nucleus made out of some critical number of TFs is mandatory for a stable propagation of protein-protein interactions as a cooperative effect. This warrants introduction of an additional nucleation step in the recruitment pathway. Typical example [36] for such mechanisms is the hybridization of complementary single stranded DNA polymers. At present we do not have enough experimental evidence to support such nucleation step in the recruitment pathway associated with the combinatorial binding of TFs with DNA.

We have considered the standard two step 3D1D model as the basic one to understand the mechanism of combinatorial binding of TFs with DNA. According to this model [16-18], TFs first nonspecifically bind with DNA and then search for their cognate sites via a combination of 1D and 3D diffusion. The 3D diffusion controlled Smolochowski rate limit seems to be enhanced by various facilitating processes viz. sliding, hopping and intersegmental transfers [16, 17]. Using the concept of hop size $k$ of our model one can generalize these processes. Particularly the non-specifically bound TFs diffuse along DNA with unit bps hop size in sliding, few bps in hopping and few hundred to thousand bps in intersegmental transfers. Here intersegmental transfers occur when two distal segments of the same DNA chain come into contact through ring-closure events over 3D space.

All these facilitating processes occur well within the Onsager radius associated with the overall electrostatic interactions at the DNA-protein interface [20]. The conformational state of DNA seems to play critical role in modulating various 1D facilitating processes. Especially sliding and hopping types of dynamics will be the dominating modes on the relaxed conformational state of DNA. Intersegmental transfer dynamics occurs mostly on the condensed state of DNA.

Detailed studies suggest that the nonspecific and site-specific binding of TFs are influenced by several factors viz. conformational state of DNA [20, 37-39], electrostatic attractive forces acting at the DNA-protein interface [40-43] and the counteracting shielding effects of solvent ions [18], presence of semi-stationary roadblocks [44] such as nucleosomes [44] and dynamic roadblocks on DNA [7, 24, 30], conformational fluctuations in the DNA binding domains of TFs [45, 46], and randomly occurring kinetic traps along the DNA sequence [8, 47-49]. Apart from these factors, the spatial organization of the genome structure also play important roles in accelerating the search process of TFs for their cognate sites on DNA [50-52]. The electrostatic interactions along with the counteracting shielding effects of solvent ions creates a fluidic type environment at the DNA-protein interface for the 1D diffusion of TFs [20].

Factors such as the overall electrostatic interactions present at the DNA-protein interface and other dynamic roadblocks are general ones hence the relative efficiencies of different modes of combinatorial binding of TFs will not be much affected by them. Combinatorial binding of TFs will be significantly modulated by the nucleosome complexes present especially on the eukaryotic genomic DNA. Nearly 147 bps of genomic DNA wrap around each nucleosome and subsequently become inaccessible for the binding of most of the TFs.

Nucleosome complexes influences the overall dynamics associated with the combinatorial binding of TFs at least in two different ways viz. they (**I**) dynamically control the accessibility of CRMs and (**II**) introduce semi-stationary roadblocks across the 1D diffusion dynamics of TFs [53, 54]. Here roadblock effects over the 1D diffusion of TFs is a general one which influence all the three modes of combinatorial binding of TFs. On the other hand, nucleosomes need to be dynamically displaced by incoming TFs to expose the CRMs which in turn requires a coordinated dynamics of several TFs. In this context one can conclude that the active recruitment mechanism of combinatorial binding of TFs with cooperative effect will be the most efficient one in the presence of nucleosomes.

## 5. Conclusions

We have developed a theoretical framework on the mechanisms of combinatorial binding of TFs which is essential for transcription regulation in eukaryotes. We have considered three possible mechanisms viz. monomer, hetero-oligomer and recruitment pathways. In the monomer pathway, combinatorial TFs search for their target sites on DNA in an independent manner and there is no protein-protein interactions among them. In the hetero-oligomer pathway, the protein-protein interactions are very strong so that the hetero-oligomer complex of TFs as a whole searches for the sequentially located cognate sites. The recruitment pathway is a special form of monomer pathway in which the TF which arrives first will recruit the adjacent TFs in a sequential manner. Under such conditions, the free energy released from the protein-protein interactions among TFs will be in turn utilized to stabilize the TFs-DNA complex. This coordinated binding of TFs in the recruitment pathway in fact emerges as a cooperative effect.

It seems that the monomer and hetero-oligomer pathways are efficient only when there are few combinatorial TFs. When the number of TFs in a combination increases then the searching efficiency of TFs in these pathways decreases with increasing number of TFs in a power law manner. The power law exponent associated with the monomer pathway seems to be strongly dependent on the number of TFs, distance between the initial position of TFs from their CRMs and the hop size associated with the dynamics of TFs on DNA. Hop size is positively correlated with the degree of condensation or supercoiling of DNA. The searching efficiency of TFs in the coordinated recruitment pathway with cooperative effects decreases with increasing number of TFs in a logarithmic manner. Since the combinatorial regulation is mandatory to avoid the speed-fidelity issues related to the searching of TFs over large genomic DNA and nuclear volumes, it is no surprise to note that the recruitment pathway with cooperative effects is the universal choice for most of the eukaryotic systems.

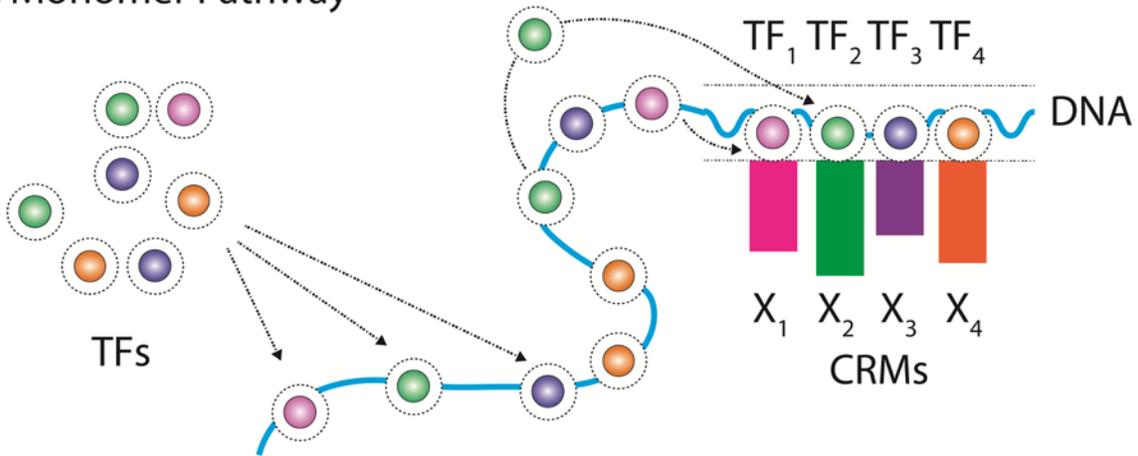

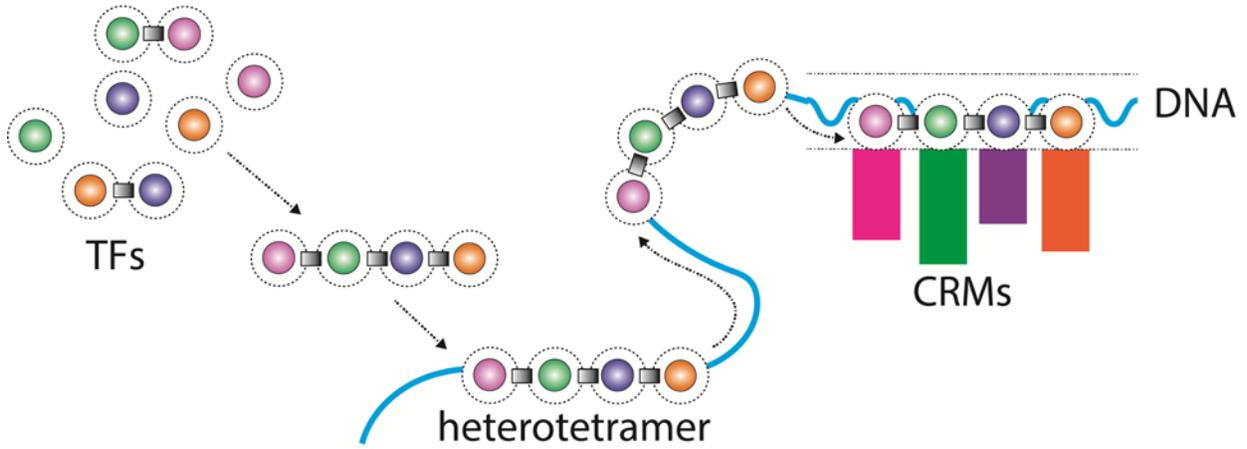

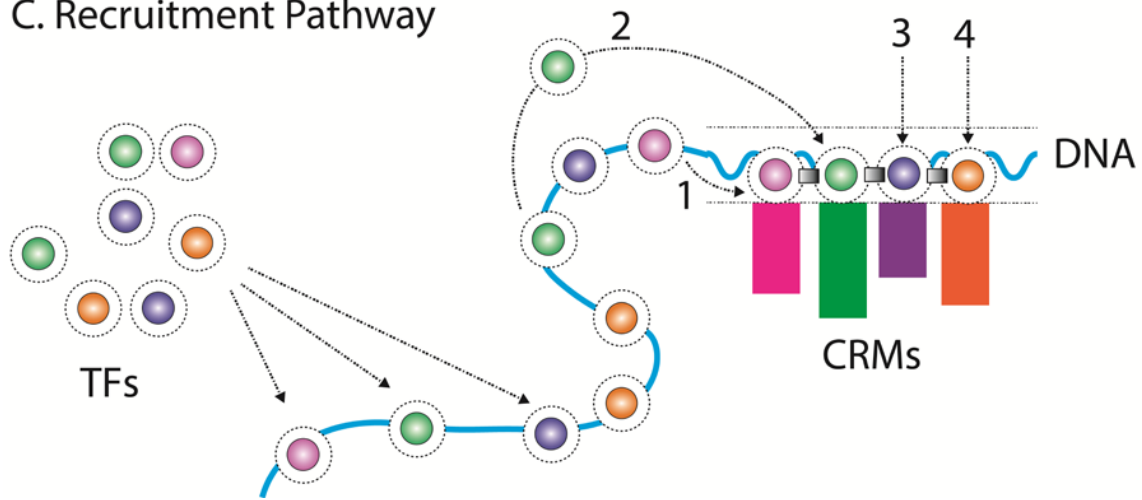

**FIGURE 1. A**. Monomer pathway. In this mode, a set of combinatorial TFs viz. $TF_1$, $TF_2$, $TF_3$ and $TF_4$ try to independently locate their sequentially located cognate sites (CRMs) $X_1X_2X_3X_4$ via a combination of 3D and 1D diffusion routes. Here the protein-protein interactions among TFs are very weak so that one can ignore them. **B**. Hetero-oligomer pathway. In this mode the protein-protein interactions among TFs are very strong so that the heterotetramer complex of the combinatorial TFs as a whole searches for the CRMs via a combination 3D and 1D diffusional routes. **C.** Recruitment pathway. In this mode, the TFs which arrives at its cognate first will in turn recruit the adjacent TFs via protein-protein interactions. Here the number 1, 2, 3, 4 indicate the order by which the respect TF arrives towards its cognate site.

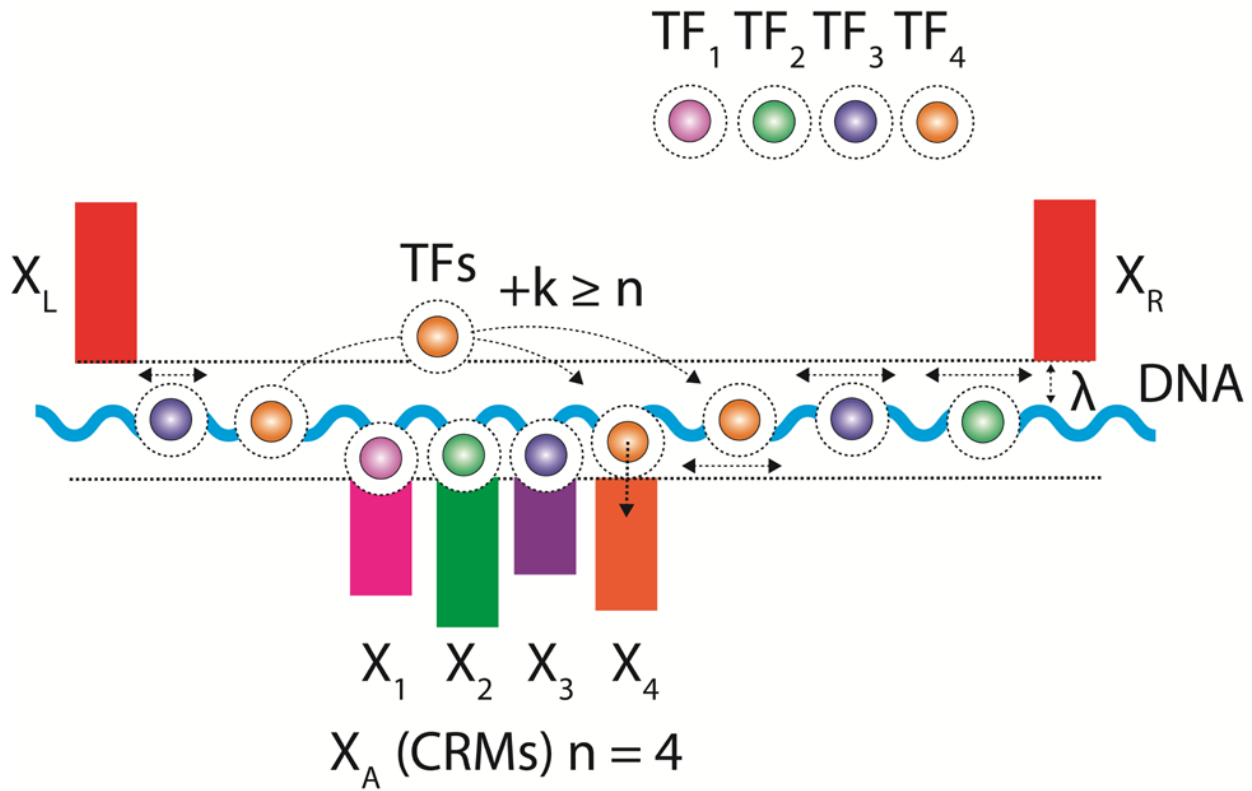

**FIGURE 2.** Various boundary conditions associated with random walk simulations. We considered a linear lattice of DNA confined in $(X_L, X_R)$ which are reflecting boundaries for the TFs = $TF_1$, $TF_2$, $TF_3$, $TF_4$ which are searching for their cognate sites located at $X_A$ = $X_1$, $X_2$, $X_3$, $X_4$. Starting from $Y_Z$ = $Y_1$, $Y_2$, $Y_3$, $Y_4$, these four TFs search for their CRM binding sites via 1D diffusion with random hops with size $k$. When $k < 4$, then a presorted initial condition is mandatory which otherwise drives the MFPT towards infinity. Here $\lambda$ is the Onsager radius [8, 20] associated with the electrostatic interactions at the DNA-protein interface. This can be defined as the distance at which the overall electrostatic interactions at the DNA-protein interface is same as that of the thermal energy ($1k_BT$).

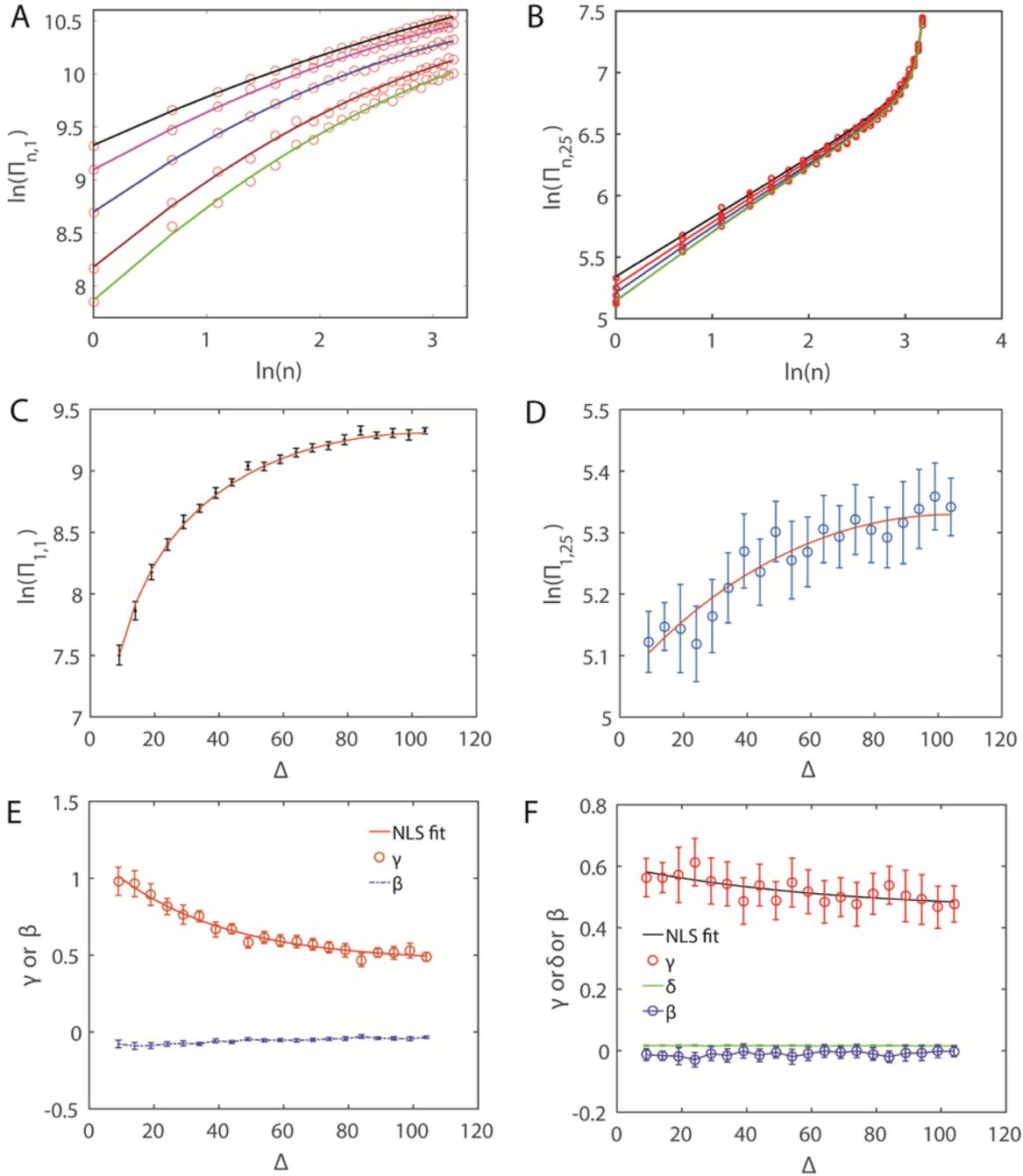

**FIGURE 3**: **A**. Stochastic random walk simulations on the binding of *n* TFs with their CRMs. Settings are $X_L = 0$, $X_R = 151$, $Y_1$ starts from 1, 50, 75, 90 and 95 to $Y_Z = Y_1, Y_2…Y_{25}$ and $X_A = X_1, X_2 … X_{25}$ starts from 105 to 130 as the number of TFs *n* increases from 1 to 25. Hop size $k = 1$. In all these simulations the mean first passage time (MFPT) that is measured in terms of dimensionless number of simulation steps was computed over $10^5$ trajectories. Red circles are simulations and solid lines $Y_1 = 1$ (black), 50 (rose), 90 (brown) and 95 (green) are the nonlinear

least square fitting with the function $\Pi_{n,1} \simeq \Pi_{1,1} n^{\gamma + \beta \ln n}$ where $\Pi_{1,1} = \left( X_1^2 - Y_1^2 \right)$ (since $D_o = 1$ for $k = 1$). **B**. Settings are $X_L = 0$, $X_R = 151$, $Y_1$ starts from 1, 50, 75, 90 and 95 to $Y_1 + 24$ and $X_A$ starts from 105 to 130 as the number of TFs $n$ increases from 1 to 25, and hop size was $k = 25$. Red circles are the stochastic simulation results and solid lines are the nonlinear least square fitting with equation $\Pi_{n,k} \simeq \Pi_{1,k} n^{\gamma + \delta/\ln(k/n) + \beta \ln n}$ where $Y_Z = 1$ (black), 50 (red), 75 (blue), 90 (brown) and 95 (green) and $\Pi_{1,k} = \left( X_1^2 - Y_1^2 \right)/D_o + X_R(1 - 1/k)$ where $D_o = (k+1)(2k+1)/6$. **C, E**. Exponents ($\gamma$, $\beta$) corresponding to nonlinear fitting in **A**. In **C** estimated values of $\Pi_{1,1}$ at various values of $\Delta = |X_1 - Y_1|$ from nonlinear least square fitting procedure is shown at the confidence level of 0.99 (black dots with error bars). Red solid line is the prediction function $\Pi_{1,1} = X_1^2 - (X_1 - \Delta)^2$ that was obtained from **Eq. 8**. In **E** though $\beta$ seems to be independent of the initial distance of TFs from the CRMs, $\gamma$ seems to be dependent on $\Delta$ in an exponential manner. Red solid line in **E** is the nonlinear least square fitting with the function $\gamma = a + b \exp(-c\Delta)$. The fitted parameters were $a = 0.46 \pm 0.05$, $b = 0.72 \pm 0.06$ and $c = 0.03 \pm 0.001$ at a confidence level of 0.95. **D, F**. Exponents ($\gamma$, $\delta$, $\beta$) corresponding to nonlinear fitting in **B**. In **D** estimated values of $\Pi_{1,25}$ at various values of $\Delta = |X_1 - Y_1|$ from nonlinear least square fitting is shown at the confidence level of 0.99 (black dots with error bars). Red solid line is the prediction function $\Pi_{1,25} \simeq 0.0045 \Pi_{1,1} + 0.96 X_R$. In case of **F** though $\beta$ and $\delta$ seem to be independent of the initial distance of TFs from the CRMs, $\gamma$ seems to be dependent on $\Delta$ in an exponential manner. Black solid line in **F** is the nonlinear least square fitting with the function $\gamma = a + b \exp(-c\Delta)$. The fitted parameters were $a = 0.45 \pm 0.1$, $b = 0.15 \pm 0.1$ and $c = 0.02 \pm 0.004$ at a confidence level of 0.95.

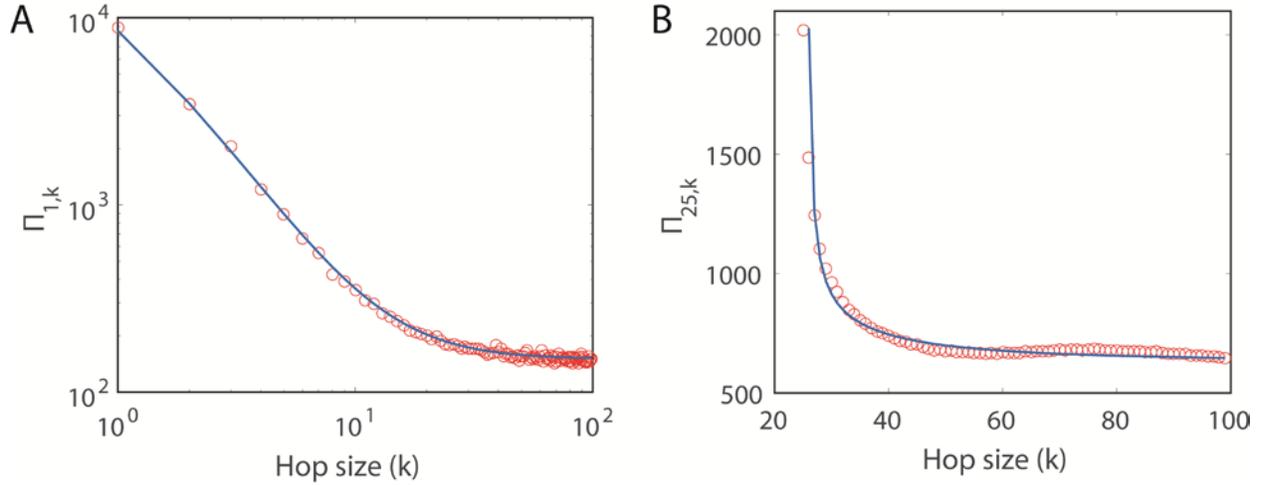

**FIGURE 4**. **A**. Effects of hop size $k$ on the overall search time associated with the binding of $n$ TFs with CRMs. Settings are $X_L = 0$, $X_R = 151$, $Y_Z = 50$ and $X_A = 105$ and $n = 1$. Hop size was iterated from $k = 1$ to 100. Hollow circles are from stochastic random walk simulations and blue solid line is the prediction from the function $\Pi_{1,k} \simeq \left[51150/(k+1)(2k+1) + 151(1-1/k)\right]$. In all these simulations the mean first passage time (MFPT) that is measured in terms of dimensionless number of simulation steps was computed over $10^5$ trajectories. **B**. Settings are $X_L = 0$, $X_R = 151$, $Y_Z = 75$ and $X_A = 105$ and $n = 25$. Hop size was iterated from $k = 26$ to 100. Hollow circles are from stochastic random walk simulations and blue solid line is the prediction from the fit function $\Pi_{25,k} \simeq \left[5400/(k+1)(2k+1) + 151(1-1/k)\right]25^{0.44+0.012/\ln(k/25)}$.